\documentclass[a4paper]{jpconf}
\usepackage{amstext}
\usepackage{amssymb}
\usepackage{subfigure}
\usepackage{color}
\usepackage{graphicx}
\usepackage{hhline}

\usepackage[colorlinks=true]{hyperref}

\def\beq{\begin{equation}}\def\eeq{\end{equation}}
\def\bea{\begin{eqnarray}}\def\eea{\end{eqnarray}}

\newfont{\cursive}{pzcmi at 9pt}
\def\msun{M_{\odot}}

\begin{document}

\title{On the distribution of stellar-sized black hole spins}

\author{Alex B. Nielsen}
\address{Max Planck Institut f\"ur Gravitationsphysik, Callinstrasse 38,
D-30167 Hannover, Germany and Leibniz Universit\"at Hannover, Welfengarten 1-A,
D-30167 Hannover, Germany}


\begin{abstract}
Black hole spin will have a large impact on searches for gravitational waves
with advanced detectors. While only a few stellar mass black hole spins have
been measured using X-ray techniques, gravitational wave detectors have the
capacity to greatly increase the statistics of black hole spin measurements. We
show what we might learn from these measurements and how the black hole spin
values are influenced by their formation channels.
\end{abstract}


\section{Introduction}
\label{sec:intro}

Gravitational wave observations will be sensitive to the spins of
black holes \cite{Cutler:1992tc}. Simulated search pipelines indicate that
accounting for black hole spin can substantially improve the sensitivity of
gravitational wave searches for binaries containing one \cite{Canton:2014ena}
or two \cite{Privitera:2013xza} black holes. However, this improvement comes at
a cost of increased computational resources and is only realised if the spins of
black holes are indeed significant. It is therefore necessary to balance the
increased cost with the expectation that black holes will indeed have
large spins. Here we discuss that evidence.

Truly isolated black holes are expected to satisfy the Kerr bound, $a_{*} \equiv
cJ/GM^2 < 1$, where $J$ and $M$ denote the angular momentum and mass of the
black hole and $c$ and $G$ the speed of light and Newton's gravitational
constant respectively. This bound is manifested in the formula for the
spatial coordinate of the event horizon in the Kerr
solution, $r=M+M\sqrt{1-a_{*}^{2}}$.

The value of $a_{*}$ can formally be computed for any object that has a mass and
angular momentum and it is worth recalling that the Kerr bound is only a bound
on black holes. Many non-compact objects such as the Earth, the Sun and
extremely rapidly rotating massive stars like VFTS 102\cite{Dufton:2011db} do
not satisfy this bound, whilst compact objects like the rapidly rotating
millisecond pulsar PSR J1748-2446ad \cite{Hessels:2006ze} and the near-extremal
black hole candidate Cygnus X-1\cite{Gou:2011nq} do. This is described in Table
\ref{Table:spins} where the spin values for the solid objects assume
they are constant density spheres.

While the exterior spacetimes of these objects are all approximately vacuum,
axisymmetric and stationary as required by the Kerr solution, except for Cygnus
X-1 they do not describe black holes (do not contain horizons) and so can have a
different multipole structure to the Kerr solution and are not constrained by
the Kerr bound. 

Things are even more extreme for elementary particles. The quantum spin of an
elementary particle can be related to an asymptotic classical angular momentum
in the sense of the Einstein-de Haas effect. In this way we can simply
calculate values for the mass, specific angular momentum and charge of
elementary particles. An electron's mass is $2.3\times 10^{-66}$secs and its
specific angular momentum is $6.4\times 10^{-22}$ so its $a_{*}$ value is a
whopping $2.8\times 10^{44}$. (In fact things are even more extreme in the
context of the Kerr-Newman solution as the electron has a square root charge of
$4.6\times 10^{-45}$ seconds). The reason for the violation of the Kerr bound in
these cases is that elementary particles typically have Planckian values of
spin, but not of mass. Of the Standard Model elementary particles, only the
Higg's boson satisfies the Kerr bound because its charge and spin are thought to
be zero.
\begin{center}
\begin{tabular}{|l|l|l|l|l|} \hline
Object & Mass [s] & J/M [s] & $a_{*}$ \\ \hhline{|=|=|=|=|}
Earth & $1.5\times 10^{-11}$ & $1.3\times 10^{-8}$ & $895$ \\ \hline
Sun & $4.9\times 10^{-6}$ & $6.1\times 10^{-6}$ & $1.2$ \\ \hline
VFTS 102 & $1.2\times 10^{-4}$ & $9.3\times 10^{-3}$ & $75$
\\ \hline
{\small PSR J1748-2446ad} & $6.9\times 10^{-6}$ & $2.9\times 10^{-6}$ &
$0.4$ \\ \hline
Cygnus X-1 & $7.30\times 10^{-5}$ & $7.23\times 10^{-5}$ &
$0.99$ \\ \hline
\end{tabular}
\label{Table:spins}
\end{center}
\textbf{Table 1.} Approximate values of mass and specific angular momentum for
the
Earth, Sun, a rapidly spinning massive star VFTS 102, a rapidly spinning
neutron star PSR J1748-2446ad and a rapidly spinning black hole Cygnus
X-1. For ease of comparison, both the mass and specific angular momentum values
are given in seconds.

\section{X-ray observations}

\begin{figure}
\centering
\begin{minipage}{.45\textwidth}
  \centering
  \includegraphics[width=1.0\linewidth]{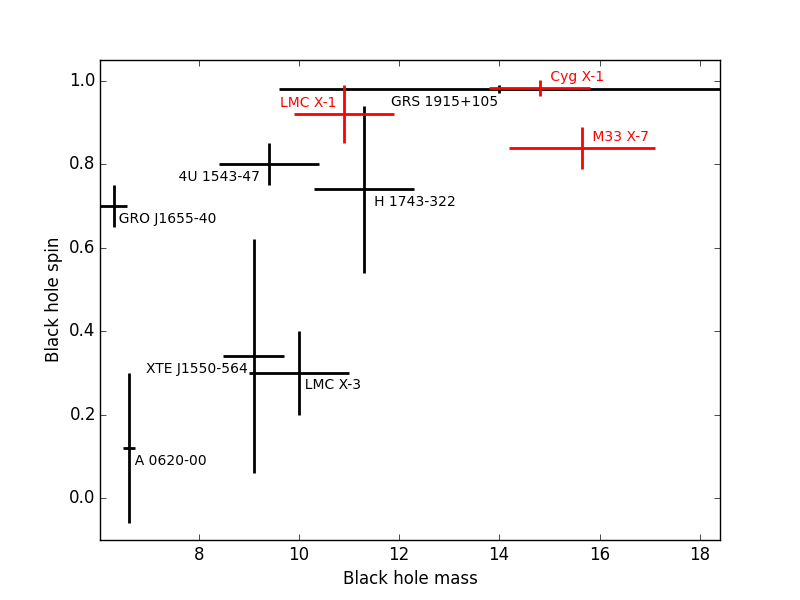}
  \caption{{\small Masses and spins for 10 black holes with approximate error
bars. The three high-mass-X-ray-binary systems, LMC X-1, Cygnus X-1 and M33 X-7
are indicated by names above the line and in red online.}}
  \label{fig:Xray_spins_names.png}
\end{minipage}
\begin{minipage}{.15\textwidth}
\end{minipage}
\begin{minipage}{.45\textwidth}
  \centering
  \includegraphics[width=1.0\linewidth]{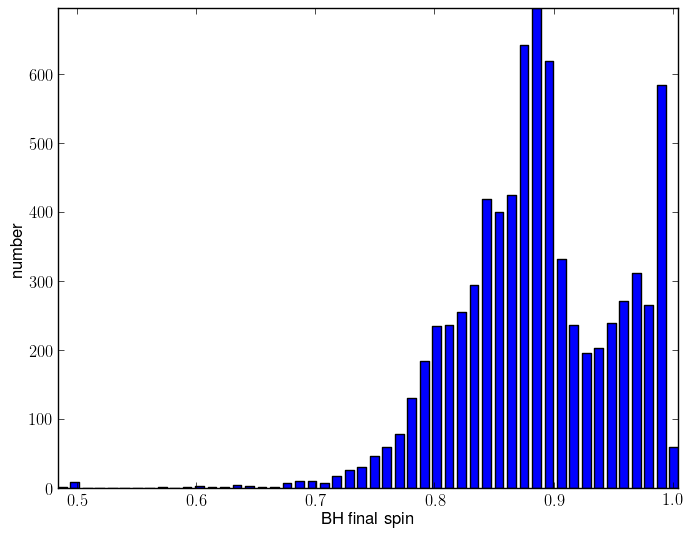}
  \caption{{\small Distribution (upper bound) of spins for black holes in
NSBH binaries, with masses greater than $2\msun$, assuming NS natal spins of 0,
BH natal spins of 0.5 and merging in 15 billion years. Adapted from
\cite{O'Shaughnessy:2005qc}.}}
  \label{fig:hist_finalspinNS0BH0pt5.png}
\end{minipage}
\end{figure}

X-ray observations of accretion disks have been able to measure the spins of
 around 10 stellar mass black holes \cite{McClintock:2011zq}. These are
displayed in Fig \ref{fig:Xray_spins_names.png}.

Of these most are Low Mass X-ray Binary (LMXB)
sources and are unlikely to form the double compact object systems necessary to
be seen by the current generation of ground-based gravitational wave detectors.
Three of the systems (LMC X-3, M33 X-7 and Cygnus X-1) are High Mass X-ray
Binaries (HMXB) where the companion to the black hole is a massive star, with
mass greater than $\sim 10\msun$ and these systems do have a chance to form
either neutron star-black hole binaries or binary black holes systems. These
systems in themselves are not targets for current ground-based GW observatories
as they are many millions of years away from merging but it is interesting to
note that all three of these HMXBs have black holes with large spins $a*>0.85$
and this suggests that the population of black holes in compact binary systems
might be dominated by black holes with large spins. The probability of obtaining
three values all above $0.7$ from a flat distribution is only $3$\% .

On the other hand, observations of neutron star pulsars suggest that the
spins of neutron stars in compact binaries are very low $a* <0.05$. If this
is indeed the case it implies that the formation channels of these objects are
very different in terms of their angular momentum evolution. Neutron stars may
be born with low spins or they may spin down rapidly after birth. The
observation of young slowly spinning pulsars, such as the Crab pulsar rotating
at only $\sim 30$Hz with a spin down rate of $\sim 0.01$Hz per year, 1000 years
after its birth, constrains models of rapid spin-down for neutron stars.

\section{Before collapse - tidal locking}

To form a compact binary system requires two large heavy stars in a
binary. Single massive stars are typically slowly
rotating at the end of their lives dues to stellar wind losses and not
differentially rotating due to angular momentum redistribution by magnetic
torques \cite{Abdikamalov:2013sta}. But stars in binaries can spin up,
increasing their rotational angular momentum by mass transfer
\cite{deMink:2012xx}. The spin rate of both stars can be affected by tides and
mass transfer \cite{deMink:2012xx}. Tidal interactions will tend to circularize
the orbits of massive stars\cite{Hurley:2002rf}. Massive stars in binaries can
have rotational velocities $\sim 200$km/s \cite{deMink:2012xx}.

Close binary systems are expected to be tidally locked \cite{Lang}. The tidal
locking timescale is between 100 and 10,000 years for helium core stars
before they turn into black holes \cite{Mendez:2010gw}. The tidal locking is
more efficient on the larger object the closer to equal mass the bodies,
perhaps giving an explanation why BHs in LMXBs have a wide range of spins,
but predominantly high spins in HMXBs. It has been shown that there is a
correlation between the black hole mass and orbital period in X-ray transients
that may be partly explained by tidal locking \cite{Lee:2001xw}. For simplicity
here we will assume that the moment of
inertia for each star is given by $kmr^{2}$ where $k$ is a constant (equal to
$2/5$ for solid balls of uniform density). The relationship between radius and
mass is often assumed to be $r \sim m^{\alpha}$ where typical values of $\alpha$
for massive stars are $0.5$ to $0.8$. Then we find that
the dimensionless spin parameter $a_{*}$ for an object with mass $m_{1}$ is
given by
\beq a_{*} = k\sqrt{\frac{c^2 r_{\odot}}{Gm_{\odot}}}
\left(\frac{r_{\odot}}{a}\right)^{3/2}\left(\frac{m_{1}}{m_{\odot}}\right)^{
2\alpha
- 1}\sqrt{\frac{m_{1}}{m_{\odot}}+\frac{m_{2}}{m_{\odot}}}.
\eeq
where $a$ denotes the semi-major axis of the orbit. Note that the three HMXB
systems would have $a_{*}>1$ if they were tidally locked stars at their current
mass and separation values.

Observed rotation rates of massive stars in binaries tend to be
significantly higher than the tidal locking rate \cite{deMink:2012xx}, since a
$10-10$ binary would need a separation of $~10r_{\odot}$ to be tidally locked
if the rotational velocity is $\sim 300$km/s, corresponding to $a_{*}\sim 100$.
Stars that are further apart will have a lower rotational velocity. A separation
of 1000 solar radii would have a tidally locked velocity of $\sim 300$m/s, since
$a_{*} \sim a^{-3/2}$.

\section{During collapse - core collapse and supernovae}

Different scenarios have been proposed for how collapse proceeds for massive
stars. These distinguish between failed core collapse (collapsar) with no
explosion and successful core collapse (supernova) where the outer envelope is
partially ejected \cite{Woosley:1993wj}. Direct collapse may be favoured
for heavy systems as the energy flux from the core is not enough to expel
the outer shell. If a star collapses directly to a black hole, without shedding
any matter and conserving its mass and its angular momentum, then its $a_{*}$
value would be conserved.

For failed core collapse models, O'Connor and Ott \cite{O'Connor:2010tk} find a
linear relationship between initial specific angular momentum of the star and
final black hole spin parameter (their fig 11). The relation is approximately
$j=2.75a_{*}$. The specific angular momentum before collapse, $j$, can be
calculated using the tidal lock model above, to obtain a final $a_{*}$ after
collapse of,
\beq a_{*} =
155k\left(\frac{r_{\odot}}{a}\right)^{3/2}\left(\frac{m}{m_{\odot}}\right)^{
2\alpha + 0.5}.
\eeq
It should be noted that for rapidly spinning black holes, centrifugal support
will tend to prevent the black hole from accreting large amounts of matter
during a collapse scenario. If an initial, near extremal
black hole forms, subsequent shells with $a_{*}>1$ will not be accreted but
expelled \cite{Lee:2001xw}. In this case, as in the successful core collapse,
detailed modeling is required to derive final mass and spin values for the
remnant.

\section{After collapse - common envelope phase}

The distribution of upper limits on the final BH spins, assuming natal spins of
zero for neutron stars and $0.5$ for black holes, for merging NSBH systems in
\cite{O'Shaughnessy:2005qc} is shown in Fig.
\ref{fig:hist_finalspinNS0BH0pt5.png}. This shows that if accretion in the
common envelope phase does behave as a thin disk, then the final
spins of BHs in NSBH binaries may be quite large, with most values falling in
the range $0.8$ to $1.0$.

The distribution in Fig. \ref{fig:hist_finalspinNS0BH0pt5.png} actually contains
two distinct populations with different formation channels. Most of the high
spin systems are actually light black holes that formed initially after collapse
as neutron stars and subsequently accreted sufficient material during common
envelope to form black holes, where the threshold for black hole formation is
set at $2\msun$. For these light systems, the accretion of a small amount of
matter is enough to spin them up considerably.

The other distribution consists of objects that collapsed to black holes before
the common envelope phase. These systems are typically more massive and hence
do not spin up as much during common envelope.

\section{Conclusion}
The limited X-ray evidence suggests that black holes in compact binaries may
have high spins. We have discussed several mechanisms that could lead to these
high spins, including tidal locking of massive progenitor stars, rapid collapse
to black holes and common envelope accretion. There are still large
uncertainties of all these processes, although they tend to reinforce one
another. Measuring the spins of a large number of black holes with
gravitational wave detectors will help to constrain some of these model
uncertainties.

\ack
  This work was supported by the Max Planck Society. The author would like to
thank Richard O'Shaughnessy for helpful conversations and sharing data for Fig.
 \ref{fig:hist_finalspinNS0BH0pt5.png}.

\end{document}